\documentstyle[12pt]{article}
\newcommand{\tev}{\,{\rm TeV}}
\title{ \vspace*{-10mm}
\begin{tabular}{ll}
\hspace*{8cm} & {\normalsize KEK-TH-630} \vspace{-3mm} \\
\hspace*{8cm} & {\normalsize May 1999}
\end{tabular} \vspace{18mm} \\
\bf Inflation in the five-dimensional universe 
with an orbifold extra dimension} 
\author{Takeshi Nihei\\
 \mbox{} \vspace{3mm}\\
 \normalsize \em  Theory Group, KEK, Tsukuba, Ibaraki 305-0801, Japan }
\date{ }
\begin{document}
\baselineskip 6mm
\renewcommand{\thefootnote}{\fnsymbol{footnote}}
\begin{titlepage}
\maketitle
\thispagestyle{empty}
\begin{abstract}
\normalsize
\baselineskip 6mm
New inflationary solutions to the Einstein equation are explicitly 
constructed in a simple five-dimensional model with an orbifold 
extra dimension $S^1/Z_2$.
We consider inflation caused by cosmological constants for 
the five-dimensional bulk and the four-dimensional boundaries.
In our solutions the extra dimension is static, and the background 
metric has a non-trivial configuration in the extra dimension. 
In addition to the solution for a vanishing bulk cosmological 
constant, which has already been discussed, 
we obtain solutions for a non-zero one. 
\end{abstract}
\end{titlepage}

\newpage
%
%
%

Recently, some people proposed a scenario to solve the hierarchy problem
by invoking large extra dimensions\cite{large-extra-dim}. 
In this scenario a fundamental Planck mass is near $\tev$, 
and the weakness of the ordinary gravity is explained by 
a large ratio between the size of the sub-millimeter extra dimensions 
and the fundamental scale. 
The extra dimensions are compactified on a manifold like $S^2$,
and the gravity can propagate in the higher-dimensional bulk, while 
the standard model fields are localized in the four-dimensional wall
with $\tev$ scale thickness in the extra dimensions. 
The extra dimensions are not homogeneous, 
since they would have special properties at the wall. 
Thus it seems important to discuss taking account of a non-trivial 
structure of the extra dimensions. 


More recently, the authors of Ref.\cite{Randall-Sundrum} proposed
a different scenario where the size of an extra dimension is 
not so large as in Ref.\cite{large-extra-dim} but the hierarchy 
is still explained by an exponential suppression factor. 
In this scenario the extra dimension is an orbifold $S^1/Z_2$. 
As in Ref.\cite{large-extra-dim}, the gravity can propagate 
in the bulk whereas the standard model fields are confined in one boundary. 
The background metric has exponential dependence in the orbifold 
direction due to non-trivial boundary conditions, 
and this behavior is essential for solving the hierarchy problem
in Ref.\cite{Randall-Sundrum}. 
This set-up is similar to that of the heterotic M-theory\cite{M-theory}, 
where the low energy effective theory of the strongly coupled heterotic 
string theory is described by an eleven-dimensional supergravity 
compactified on $S^1/Z_2$, and super Yang-Mills gauge multiplets
live on the two boundaries. 
A non-trivial metric distortion in the orbifold dimension is 
one of the important ingredients in the M-theory too. 


The extra dimensions mentioned above 
might play an important role in cosmology. 
Inflationary solutions of the Einstein equation and 
early cosmology in models with extra dimensions have been discussed 
by many authors. 
In some analyses, the metric is assumed to be uniform in 
the extra dimension\cite{large-extra-dim-inflation}\cite{trivial-metric}.
If the metric has non-trivial dependence in the extra 
dimension at the early universe, the analyses on inflation may be altered. 
There also analyses where non-trivial background metrics
are considered\cite{M-inflation}\cite{dilatonic-domain-wall}\cite{Benakli}\cite{brane-inflation}\cite{Kaloper-Linde}\cite{Binetruy}. 
However inflationary exact solutions in the presence of both a bulk 
cosmological constant and boundary ones have not been discussed 
in these papers. 


In this paper, we try to find new inflationary solutions with 
non-trivial configurations in the extra dimension.
We restrict ourselves to a five-dimensional gravity with an orbifold 
extra dimension $S^1/Z_2$. 
Starting from a five-dimensional Einstein action with two boundary terms, 
we analyze classical solutions to the Einstein equation. 
The five-dimensional bulk action is described 
by a five-dimensional gravity and a cosmological constant. 
For the boundary actions, we treat them as localized cosmological
constants. 
With simple assumptions for the metric, 
we find new inflationary solutions which have 
non-trivial dependences in the extra dimension. 
In addition to the solution for a vanishing bulk cosmological 
constant, which has already been discussed in Ref.\cite{M-inflation}, 
we obtain solutions for a non-zero one.

\vspace{1cm} 	
%
%
%

We consider the five-dimensional universe compactified on an 
orbifold $S^1/Z_2$. 
We use coordinates $x^M$ $=$ $(x^\mu, y)$, 
where $M$ $=$ $0,1,2,3,5$ is an index for the five-dimensional space,
and $\mu$ $=$ $0,1,2,3$ is that for the uncompactified four-dimensional space. 
The coordinate $y$ is assigned to the extra orbifold dimension
with an identification $y+2L$ $\sim$ $y$. 
To describe $S^1/Z_2$, we work on orbifold picture. 
Namely, we analyze physics in the region $-L$ $\leq$ $y$ $\leq$ $L$, 
requiring that every field is even under $Z_2$ action 
$y$ $\rightarrow$ $-y$.
Fixed points of the $Z_2$ action are located at $y$ $=$ 0, $L$.
These two points correspond to two boundaries of the 
segment $S^1/Z_2$. 


The action is given by
\begin{eqnarray}
S & = & -\frac{1}{2 \kappa_5^2} \int_{-L}^{L} d y \int d^4 x
\sqrt{g} (R+2 \Lambda) 
        + \int d^4 x \sqrt{-g_1} {\cal L}_1
        + \int d^4 x \sqrt{-g_2} {\cal L}_2.
\label{eqn:action}
\end{eqnarray}
Here the first term represents the five-dimensional bulk action 
which includes the five-dimensional gravitational fields and 
the cosmological constant $\Lambda$. 
The second and the third terms are 
the four-dimensional boundary actions localized at 
$y$ $=$ 0 and $y$ $=$ $L$, respectively.
The fields in the standard model are supposed to be confined 
at one boundary,  and some hidden matter fields are confined at 
the other boundary. 
For simplicity, we assume that the boundary potentials
have slow-roll properties, 
and we can treat these boundary terms as localized
cosmological constants during inflation. 
The $g_1$ and $g_2$ are metrics on the boundaries, 
and are written by the five-dimensional metric $g_{MN}$ as
$g_1^{\mu \nu}$ $=$ $g^{\mu \nu}(y=0)$ and 
$g_2^{\mu \nu}$ $=$ $g^{\mu \nu}(y=L)$. 
The sign convention for the metric is $(+,-,-,-,-)$. 
Note that the physical length of the orbifold dimension
is $L_{\rm phys}$ $=$ $\int dy \sqrt{-g_{55}}$. 
The five-dimensional gravitational coupling constant $\kappa_5$ 
is related to the four-dimensional Newton constant $G_N$ 
as $\kappa_5^2$ $=$ $16 \pi G_N L_{\rm phys}$. 

Minimizing the action (\ref{eqn:action}), 
we obtain the Einstein equation as\cite{Randall-Sundrum}
\begin{eqnarray}
\lefteqn{ \sqrt{g}\left(R^{MN}-\frac{1}{2}g^{MN}R \right) } \nonumber \\
& = & - \kappa_5^2 \left[ 
\sqrt{-g_1} g_1^{\mu \nu} \delta^M_{\mu} \delta^N_{\nu}{\cal L}_1 \delta(y)
+ \sqrt{-g_2} g_2^{\mu \nu} \delta^M_{\mu} 
\delta^N_{\nu}{\cal L}_2 \delta(y-L) \right]
+ \sqrt{g}g^{MN}\Lambda. 
\label{eqn:Einstein-eq}
\end{eqnarray}
In deriving Eq.(\ref{eqn:Einstein-eq}), we neglect dynamics of the 
fields on the boundaries. 
Namely, ${\cal L}_1$ and ${\cal L}_2$ in this equation are treated 
as constants. 
We put the following ansatz for the metric: 
\begin{eqnarray}
ds^2 = g_{MN} dx^M dx^N = 
u(y)^2 dt^2 - a(y,t)^2 d \vec{x}^2 - b(y,t)^2 d y^2, 
\label{eqn:metric}
\end{eqnarray}
where $u$, $a$ and $b$ are scale factors for $t$ $( = x^0)$, $x^{1, 2, 3}$ 
and $y$, respectively. 
With this metric, Eq.(\ref{eqn:Einstein-eq}) reduces to 
the following relations:
\begin{eqnarray}
\frac{1}{u^2} \left[ \left(\frac{\dot{a}}{a}\right)^2
+ \frac{\dot{a}}{a}\frac{\dot{b}}{b} \right]
- \frac{1}{b^2} \left[ \frac{a''}{a}
+ \left(\frac{a'}{a}\right)^2
- \frac{a'}{a}\frac{b'}{b} \right]
 & = & - \frac{\kappa_5^2}{3b}
\left[ \delta (y) {\cal L}_1 
       +\delta (y-L) {\cal L}_2 \right] + \frac{\Lambda}{3}, 
\label{eqn:R00}\\
\lefteqn{ \hspace*{-8.0cm}
\frac{1}{u^2} \left[ 2 \frac{\ddot{a}}{a}
+ \frac{\ddot{b}}{b} + \left(\frac{\dot{a}}{a}\right)^2
+ 2 \frac{\dot{a}}{a}\frac{\dot{b}}{b} \right]
- \frac{1}{b^2} \left[ 2 \frac{a''}{a} 
+ \frac{u''}{u} + \left(\frac{a'}{a}\right)^2
+ 2 \frac{u'}{u}\frac{a'}{a}
- 2 \frac{a'}{a}\frac{b'}{b}
- \frac{u'}{u}\frac{b'}{b} \right]     } \nonumber \\
 & = & - \frac{\kappa_5^2}{b} 
\left[ \delta (y) {\cal L}_1 
       +\delta (y-L) {\cal L}_2 \right] + \Lambda, 
\label{eqn:Rmn}\\
\frac{1}{u^2} \left[ \frac{\ddot{a}}{a}
+ \left(\frac{\dot{a}}{a}\right)^2 \right]
- \frac{1}{b^2} \left[ \left(\frac{a'}{a}\right)^2
+ \frac{u'}{u}\frac{a'}{a} \right]
 & = & \frac{\Lambda}{3}, 
\label{eqn:R55}\\
-\frac{\dot{a}'}{a} + \frac{u'}{u} \frac{\dot{a}}{a}
+ \frac{a'}{a} \frac{\dot{b}}{b} & = & 0, 
\label{eqn:R05}
\end{eqnarray}
where primes and dots denote derivatives with respect to 
$y$ and $t$, respectively. 
We look for classical solutions to these equations from now on. 


We focus on Eq.(\ref{eqn:R05}) first. 
Useful observation follows if we assume that 
$u(y)$ and $a(y,t)$ have common $y$-dependence 
and $a(y,t)$ is separable:
\begin{eqnarray}
u(y) = f(y), \ \ \ a(y,t) = f(y)v(t).
\label{eqn:ansatz}
\end{eqnarray}
With this assumption, the first and the second terms in the left-hand
side of Eq.(\ref{eqn:R05}) cancel out, and we obtain 
$\dot{b}=0$ or $a'=0$. 
However, for the latter case, Eqs.(\ref{eqn:R00}) and (\ref{eqn:Rmn}) cannot 
be satisfied due to the non-zero boundary terms. 
Therefore it follows that the extra dimension is automatically static:
\begin{eqnarray}
\dot{b}(y,t) & = & 0. 
\label{eqn:static}
\end{eqnarray}
Note that if both ${\cal L}_1$ and ${\cal L}_2$ vanish, there is also
a possibility of $\dot{b} \neq 0$. 
Though $b(y,t)$ is time independent, it may have non-trivial $y$-dependence. 
For simplicity, we assume that $b(y,t)$ also has the same $y$-dependence
as $u(y)$ and $a(y,t)$: 
\begin{eqnarray}
b(y,t) & = & f(y). 
\label{eqn:form-of-b}
\end{eqnarray}


We can solve $v(t)$ and $f(y)$ easily. 
For $v(t)$, we obtain $\ddot{v}/v$ $=$ $(\dot{v}/v)^2$ from 
Eqs.(\ref{eqn:R00}), (\ref{eqn:Rmn}), 
(\ref{eqn:ansatz}) and (\ref{eqn:form-of-b}).
Thus the three spatial dimensions expand exponentially:
\begin{eqnarray}
v(t)& = & e^{Ht}, 
\label{eqn:inflation}
\end{eqnarray}
where $H$ is a constant,
and we choose the direction of time so that $H$ $\geq$ 0. 
In this way, assuming Eqs.(\ref{eqn:ansatz}) and (\ref{eqn:form-of-b}),
we can realize the exponential expansion of the universe together with 
the static extra dimension. 
The expansion parameter $H$ is determined later. 
Collecting everything, we have assumed the metric
\begin{eqnarray}
ds^2 & = & f(y)^2 (dt^2 - e^{2Ht}d \vec{x}^2 - d y^2). 
\label{eqn:metric-final}
\end{eqnarray}
The geometry of hypersurfaces $y$ $=$ constant is that of the
de Sitter space\cite{M-inflation}\cite{Vilenkin}. 
For $f(y)$, Eqs.(\ref{eqn:R00})-(\ref{eqn:R55}) provide 
the following equations:
\begin{eqnarray}
\left( \frac{f'}{f} \right)^2 
& = & H^2 -\frac{\Lambda}{6} f^2, 
\label{eqn:f1}\\
\frac{f''}{f} & = & \frac{\kappa_5^2}{3} f \left[\delta(y){\cal L}_1
+\delta(y-L){\cal L}_2 \right] + H^2 - \frac{\Lambda}{3}f^2.
\label{eqn:f2}
\end{eqnarray}
We require that the scale factor $f$ is continuous across the boundaries,
while we allow its derivatives to be discontinuous. 
The curvature of the five-dimensional space is constant 
$R$ $=$ $-10\Lambda/3$ except at the boundaries.
The equations (\ref{eqn:f1}) and (\ref{eqn:f2}) have different 
types of solutions for $\Lambda =0$, $\Lambda>0$ and $\Lambda<0$. 

\vspace{1cm} 	
%
%

First we describe the case $\Lambda=0$. 
The solution to Eq.(\ref{eqn:f1}) 
consistent with the orbifold symmetry 
$f(-y)$ $=$ $f(y)$ is given by
\begin{eqnarray}
f(y) & = & e^{-H |y|+c_0},
\label{eqn:solution0}
\end{eqnarray}
where $c_0$ is a constant. 
This solution corresponds to the solution in 
Refs.\cite{M-inflation} and \cite{Vilenkin}, 
and also can be written in a form similar to the solution 
in Ref.\cite{Kaloper-Linde} by a coordinate transformation $Y$ $=$ 
$[-e^{-H |y|+c_0}+1]/H$. 
In addition to Eq.(\ref{eqn:solution0}), 
a solution $e^{H|y|+c_0} $ is also possible. 
The general solution for $\Lambda$ $=$ 0 without the ansatz 
(\ref{eqn:metric-final}) is given in Ref.\cite{M-inflation}. 
Note that we use the solution (\ref{eqn:solution0}) only 
for $-L$ $<$ $y$ $\leq$ $L$, and keep in mind that 
$f(y)$ is a periodic function. 
Explicitly, we use formulas 
$|y|'$ $=$ $\epsilon(y)$ $-$ $\epsilon(y-L)$ $-1$
and $|y|''$ $=$ $2\delta(y)$ $-$ $2\delta(y-L)$, 
where $\epsilon(y)$ is 1 for $y \geq 0$ and $-1$ for $y<0$. 
Calculating $f''$ from Eq.(\ref{eqn:solution0}), 
the delta-functions $\delta(y)$ and $\delta(y-L)$ arise.
For general values of ${\cal L}_1$ and ${\cal L}_2$, 
Eq.(\ref{eqn:solution0}) is not compatible with Eq.(\ref{eqn:f2}). 
However, if we assume the following conditions, 
Eq.(\ref{eqn:solution0}) is consistent with Eq.(\ref{eqn:f2}):
\begin{eqnarray}
\frac{\kappa_5^2}{3}{\cal L}_1
 = - 2 H e^{-c_0}, \ \ \ 
\frac{\kappa_5^2}{3}{\cal L}_2
 = 2 H e^{HL-c_0}.
\label{eqn:condition0}
\end{eqnarray}
We must require ${\cal L}_1$ $<$ 0 and ${\cal L}_2$ $>$ 0
to satisfy these conditions for $H$ $>$ 0. 
We must also require ${\cal L}_1$ $+$ ${\cal L}_2$ $>$ 0,
since the physical length of the extra dimension
$L_{\rm phys}$ $=$ $\int dy \sqrt{-g_{55}}$ 
$=$ $-(6/{\cal L}_1+6/{\cal L}_2)/\kappa_5^2$
must be positive. 
For ${\cal L}_1$ $>$ 0 and ${\cal L}_2$ $<$ 0, we can satisfy
similar conditions with another solution $e^{H|y|+c_0}$. 
In the case of ${\cal L}_1 {\cal L}_2$ $>$ 0, 
we have no solution for $\Lambda$ $=$ 0 
under the present ansatz for the metric (\ref{eqn:metric-final}).
We can determine $H$ and $c_0$ from Eq.(\ref{eqn:condition0}) as 
$H$ $=$ $\log ( - {\cal L}_1/{\cal L}_2 )$ $/L$ and 
$c_0$ $=$ $\log ( -6H/\kappa_5^2{\cal L}_1)$. 
For ${\cal L}_1$ $=$ ${\cal L}_2$ $=$ 0, which corresponds to $H$ $=$ 0, 
the solution (\ref{eqn:solution0}) reduces to $f$ $=$ constant.
Then the metric is static and uniform in the extra dimension. 
Note that for vanishing ${\cal L}_1$ and ${\cal L}_2$ we also have 
a possibility $a'$ $=0$ instead of Eq.(\ref{eqn:static}). 


The expansion rate depends on the position 
in the extra dimension\cite{M-inflation}.
After we canonically normalize the four-dimensional coordinate so 
that $ds^2$ $=$ $dt^2$ $-$ $v^2 d \vec{x}^2$ for a fixed $y$, 
we observe that 
in general the effective expansion rate $H_{\rm eff}(y)$ 
$\equiv$ $H/f(y)$ depends
on the position in the extra dimension: 
$H_{\rm eff}(y)$ $=$ 
$(-\kappa_5^2{\cal L}_1/6)$ $( -{\cal L}_1/{\cal L}_2)^{|y|/L}$.
In particular the expansion rates for the boundaries are given by 
$H_{\rm eff}(0)$ $=$ $-\kappa_5^2{\cal L}_1/6$ and 
$H_{\rm eff}(L)$ $=$ $\kappa_5^2{\cal L}_2/6$. 


The relation between the expansion rate and the cosmological constant 
mentioned above 
is quite unconventional as pointed out in Ref.\cite{M-inflation}. 
In a usual four-dimensional scenario, 
we have a relation $H$ $\sim$ $\sqrt{V}/M_{\rm Pl}$, 
where $V$ denotes an inflaton potential and $M_{\rm Pl}$ is 
the four-dimensional Planck mass. 
On the other hand, in the present case we have $H$ $\sim$ 
$-\kappa_5^2 {\cal L}_1$ $\sim$ $L_{\rm phys}V/M_{\rm Pl}^2$,
where we put ${\cal L}_1$ $\sim$ $-V$. 
The expansion rate is suppressed by $M_{\rm Pl}^2$ rather than
$M_{\rm Pl}$, hence it may be difficult to obtain a sufficient $e$-folding
unless we take $L_{\rm phys}$ $\sim$ $1/M_{\rm Pl}$. 


We can also find a solution for $\Lambda > 0$. 
In this case, the solution to Eq.(\ref{eqn:f1}) 
consistent with the orbifold symmetry is
\begin{eqnarray}
f(y) & = & \sqrt{\frac{6 H^2}{\Lambda}} 
\frac{1}{\cosh ( H|y|+ c_+ )},
\label{eqn:solutionP}
\end{eqnarray}
where $c_+$ is a constant. 
Conditions for Eq.(\ref{eqn:solutionP}) to be consistent 
with Eq.(\ref{eqn:f2}) are
\begin{eqnarray}
\frac{\kappa_5^2}{3}{\cal L}_1
 & = & -2 \sqrt{\frac{\Lambda}{6}} \sinh (c_+), \nonumber \\ 
\frac{\kappa_5^2}{3}{\cal L}_2
 & = & 2 \sqrt{\frac{\Lambda}{6}} \sinh ( HL + c_+ ).
\label{eqn:conditionP}
\end{eqnarray}
From these relations, 
we can determine $H$ and $c_+$ 
in terms of ${\cal L}_1$, ${\cal L}_2$ and $\Lambda$. 
The physical length of the extra dimension is given by
$L_{\rm phys}$ $=$ $\sqrt{6/\Lambda}$ ($\arctan \theta_1$ $+$ 
$\arctan \theta_2$), where $\theta_i$ $=$ $\sqrt{6/\Lambda}$
$\kappa_5^2 {\cal L}_i/6$. 
Hence we must require ${\cal L}_1$ $+$ ${\cal L}_2$ $>0$ to ensure
$L_{\rm phys}$ $>0$. 
The shape of the solution (\ref{eqn:solutionP}) depends on the 
signatures of the boundary terms ${\cal L}_1$ and ${\cal L}_2$. 
For ${\cal L}_1$ $<$ 0 and ${\cal L}_2$ $>$ 0, 
we have a solution with $c_+$ $>$ 0. 
In this case the shape of the solution in the region 0 $<$ $y$ $<$ $L$ 
is similar to that of the case $\Lambda$ $=$ 0.
The scale factor $f$ has a maximum at $y$ $=$ 0 and a minimum
at $y$ $=$ $L$. 
For ${\cal L}_1$ $>$ 0 and ${\cal L}_2$ $>$ 0, 
we have a solution with $-HL$ $<$ $c_+$ $<$ 0. 
In this case the scale factor has a maximum at $y$ $=$ $-c_+/H$ and 
a minimum at $y$ $=$ 0 or $y$ $=$ $L$. 
For ${\cal L}_1$ $>$ 0 and ${\cal L}_2$ $<$ 0, 
the scale factor has a maximum at $y$ $=$ $L$ and a minimum
at $y$ $=$ 0. 
We have no solution for ${\cal L}_1$ $<$ 0 and ${\cal L}_2$ $<$ 0. 

The effective expansion rate $H_{\rm eff}(y)$ depends on 
${\cal L}_1$, ${\cal L}_2$ and $\Lambda$. 
If the bulk contribution $\sqrt{\Lambda}$ is much larger than 
the magnitude of the boundary terms $\kappa_5^2{\cal L}_1$ and 
$\kappa_5^2{\cal L}_2$, 
the effective expansion rate is dominated by $\Lambda$ in the 
whole region of the orbifold: $H_{\rm eff}(y)$ $\sim$ $\sqrt{\Lambda/6}$. 
If $\Lambda$ is much smaller than the boundary contributions, 
$H_{\rm eff}(y)$ reduces to the result in the case $\Lambda$ $=$ 0: 
$H_{\rm eff}(0)$ $=$ $|\kappa_5^2{\cal L}_1/6|$ and 
$H_{\rm eff}(L)$ $=$ $|\kappa_5^2{\cal L}_2/6|$.
Notice that the extra dimension is static even in the 
presence of the bulk cosmological constant $\Lambda$. 
This constant works to generate the metric distortion in the extra
dimension and also to inflate the four-dimensional subspaces. 


Finally we describe the case $\Lambda < 0$. 
The solution to Eq.(\ref{eqn:f1}) is obtained as follows:
\begin{eqnarray}
f(y) & = & \sqrt{\frac{6 H^2}{-\Lambda}} 
\frac{1}{\sinh ( H|y|+ c_- )}.
\label{eqn:solutionN}
\end{eqnarray}
This scale factor is finite everywhere in the extra dimension if $c_- >0$. 
We can also find a similar solution $\sim$ $1/\sinh (-H|y|+ c_-)$ with
$c_- > HL$. 
Conditions for Eq.(\ref{eqn:solutionN}) to be consistent 
with Eq.(\ref{eqn:f2}) are written as
\begin{eqnarray}
\frac{\kappa_5^2}{3}{\cal L}_1
 & = & -2 \sqrt{\frac{-\Lambda}{6}} \cosh (c_-), \nonumber \\ 
\frac{\kappa_5^2}{3}{\cal L}_2
 & = & 2 \sqrt{\frac{-\Lambda}{6}} \cosh ( HL + c_- ).
\label{eqn:conditionN}
\end{eqnarray}
To satisfy these conditions, we must require 
${\cal L}_1$ $<$ 0 and ${\cal L}_2$ $>$ 0. 
The scale factor (\ref{eqn:solutionN}) has a maximum at $y$ $=$ 0 
and a minimum at $y$ $=$ $L$. 
For ${\cal L}_1$ $>$ 0 and ${\cal L}_2$ $<$ 0,
we can satisfy similar conditions with another solution
$\sim$ $1/\sinh (-H|y|+ c_-)$. 
We have no solution for ${\cal L}_1 {\cal L}_2$ $>$ 0. 
The physical length of the extra dimension $L_{\rm phys}$ is 
determined by ${\cal L}_1$, ${\cal L}_2$ and $\Lambda$ in a 
similar way to the cases $\Lambda$ $=0$ and $\Lambda > 0$.

\vspace{1cm} 	
%
%

There are a few comments in order. 
We can also find inflationary solutions to 
Eqs.(\ref{eqn:R00})-(\ref{eqn:R05})
by assuming $b$ $=$ constant instead of Eq.(\ref{eqn:form-of-b}). 
However, with this assumption, it follows that
we must tune the parameters
$\Lambda$, ${\cal L}_1$ and ${\cal L}_2$ to obtain solutions. 
This tuning is similar to that in 
Refs.\cite{Randall-Sundrum} and \cite{Binetruy}.
On the other hand, with the assumption (\ref{eqn:form-of-b}) 
which we adopted, we do not need to tune the parameters
to find solutions, 
and we can rather determine the expansion parameter $H$ according to 
these parameters. 
Hence we expect that the assumption (\ref{eqn:form-of-b}) is more 
likely than $b$ $=$ constant unless we introduce some principle 
to justify such a tuning of parameters. 


The general solution for $\Lambda =0$
is given in Ref.\cite{M-inflation}, 
where $u$, $a$ and $b$ in Eq.(\ref{eqn:metric}) 
are allowed to be general functions of $t$ and $y$. 
In that analysis it is discussed that the solution with the 
static extra dimension corresponds to a specific choice of initial
conditions, and all other choices of the initial conditions lead to
a collapse of the extra dimension or the three spatial dimensions. 
This implies that the solution (\ref{eqn:solution0}) for $\Lambda =0$
is unstable under a small perturbation. 
It is worth studying whether the situation is the same or not 
for $\Lambda$ $\neq 0$.


In addition to the problem stated above, there remain a lot of
works to be done in order to 
construct a realistic scenario in this framework. 
To describe the scenario, 
we must include dynamics of inflaton fields. 
For example, we replace the bulk cosmological constant $\Lambda$ 
by a Lagrangian for bulk inflaton fields, 
or take into account the time dependence of 
Lagrangians for the boundary inflaton fields
included in ${\cal L}_1$ and ${\cal L}_2$. 
We expect that the inflating universe is in a false vacuum,
and the universe go to the true vacuum in the end. 
In such a set-up, we should discuss various subjects:
slow-roll condition,
$e$-folding number, 
density fluctuation, 
end of inflation and reheating temperature, 
big-bang nucleosynthesis, and so on. 

%
\vspace{1cm}

%
%
%

In summary, we have constructed 
new inflationary solutions to the Einstein equation
in a simple five-dimensional theory with an orbifold 
extra dimension $S^1/Z_2$.
We have considered inflation caused by cosmological constants for 
the five-dimensional bulk and the four-dimensional boundaries. 
The solutions have non-trivial behaviors in the extra dimension. 
In addition to the solution for a vanishing bulk cosmological 
constant, we have obtained solutions for a non-zero one. 

%
%
%
%

We would like to thank Y. Okada for a useful discussion.
This work was supported in part by
the Grant-in-Aid for Scientific
Research from the Ministry of Education, Science and Culture, Japan. 
%

%
%
%
%
\end{document}